\documentclass[aps,twocolumn,a4paper,showpacs]{revtex4}
\usepackage{latexsym}
\usepackage{amssymb}
\usepackage{amsmath}
\usepackage{amsfonts}
\usepackage[dvipdfmx]{graphicx}
\usepackage[dvipdfmx]{color}
\usepackage{txfonts}
\usepackage{bm}

\begin{document}
\title{Heat transport in nonlinear lattices free from the Umklapp process}

\author{Kazuyuki Yoshimura$^1$, Yusuke Doi$^2$, and Tomoya Kitamura$^3$}
\affiliation{$^1$Faculty of Engineering, Tottori University,
 4-101 Koyama-Minami, Tottori 680-8552, Japan\\
 $^2$Division of Mechanical Engineering, Graduate School of Engineering\\
 Osaka University, 2-1 Yamadaoka, Suita, Osaka 565-0871, Japan\\
 $^3$Department of Information and Electronics, Graduate school of Engineering\\
 Tottori University, 4-101 Koyama-Minami, Tottori 680-8552, Japan}
\begin{abstract}
   We construct one-dimensional nonlinear lattices
 having the special property
 such that
 the Umklapp process vanishes and only the normal processes
 are included in the potential functions.
 These lattices have
 long-range quartic nonlinear
 and nearest neighbor harmonic interactions
 with/without harmonic on-site potential.
 We study heat transport in
 two cases of the lattices with and without harmonic on-site potential
 by non-equilibrium molecular dynamics simulation.
 It is shown that
 the ballistic heat transport occurs in both cases,
 i.e.,
 the scaling law $\kappa\propto N$ holds between 
 the thermal conductivity $\kappa$ and the lattice size $N$.
 This result 
 directly validates Peierls's hypothesis that
 only the Umklapp processes can cause the thermal resistance
 while the normal one do not.
\end{abstract}
%
%
\maketitle
%
%
%
%
\section{Introduction}
\label{sec:introduction}

   Heat transport is ubiquitous in nature.
 In macroscopic scale materials,
 it is well described by the Fourier's law
 $J=-\kappa\nabla T$,
 where $J$ and $\nabla T$ are
 the heat flux and the temperature gradient, respectively,
 and $\kappa$ is a constant called the thermal conductivity,
 whose value is determined by the material.
 Consider a one-dimensional shaped material with length $L$
 which is kept with different temperatures at the both ends.
 The Fourier's law implies that
 the heat flux $J$ is attenuated
 as $J\propto L^{-1}$ with increasing $L$,
 under a given temperature difference.
 This attenuation
 indicates the existence of thermal resistance. 
 In microscopic scale,
 the value of $\kappa$ depends on both the material and its length \cite{Chang-2008},
 but the heat flux is still attenuated as the length increases,
 i.e., the thermal resistance still emerges.
 It has been a long-standing unsolved problem
 to clarify
 the origin of thermal resistance
 based on the dynamics of atoms.
 
   A simple microscopic model for solids
 is one-dimensional lattice,
 and it has been used for studying heat transport
 via atomic vibrations \cite{Kittel-2004,Ashcroft-1976}.
 Nonlinearity of the lattice,
 which is necessary for the phonon interactions,
 is essential to explain the emergence of thermal resistance.
 There are two types of the phonon interaction processes,
 which are called the {\it normal} and the {\it Umklapp} processes.
 Peierls posed the hypothesis that
 only the Umklapp processes can cause the thermal resistance
 while the normal one do not \cite{Peierls-1955,Peierls-1997},
 and this hypothesis has been widely believed so far.
 However, at least in classical physics regime,
 the hypothesis does not have a firm theoretical basis.
 To the best of our knowledge,
 only the existing basis is that
 in a lattice with periodic boundary conditions
 the harmonic part of its total heat flux is conserved
 if there is no Umklapp process,
 provided that the lattice has {\it no dispersion} \cite{Peierls-1997,Jackson-1978,Lepri-2003}.
 The assumption of no dispersion is never satisfied
 in one-dimensional lattices,
 and this is not a satisfactory basis for the hypothesis.
 
   The above hypothesis has not yet been verified
 even by numerical simulations.
 The crucial reason is a lack of a nonlinear lattice model
 that is free from the Umklapp process.
 In the present paper, we construct
 a class of nonlinear lattices without the Umklapp process,
 which we call the {\it Umklapp-free lattices} (UFLs).
 The UFLs have
 long-range quartic nonlinear
 and nearest neighbor harmonic interactions with/without harmonic on-site potential.
 They closely relate with
 the {\it Pairwise Interaction Symmetric Lattice} (PISL)
 \cite{Doi-2016,Doi-2020},
 which is a special lattice model recently constructed
 and having a hidden symmetry in its potential function
 to enhance the mobility of a localized mode called
 the {\it discrete breather}
 \cite{Takeno-1988,Page-1990,Yoshimura-2021,Yoshimura-2014}.
 
   The UFL enables one to directly verify Peierls's hypothesis.
 We numerically
 study heat transport in two types of UFLs,
 which are with and without harmonic on-site potential,
 and show that
 the ballistic heat transport occurs in both of the UFLs,
 i.e., $\kappa\propto N$ holds between 
 the thermal conductivity $\kappa$ and the lattice size $N$.
 Our results justify
 Peierls's hypothesis
 at least in the present lattice models.
 
   We mention known results about heat transport in the PISL,
 as it is a model closely related to the UFL.
 A near ballistic transport,
 $\kappa\propto N^\alpha$ with $\alpha\simeq 1$,
 has been reported in some works
 \cite{Bagchi-2017,Yoshimura-2019,Yoshimura-2020,Bagchi-2021},
 whereas a different value $\alpha\simeq 0.71$ in \cite{Wang-2020}. 
 It is still unclear whether the PISL exhibits
 the ballistic transport or non-ballistic but anomalous one.
 However, at least,
 the PISL seems to have $\alpha$ significantly larger
 than nonlinear lattices which are known to exhibit anomalous heat transport
 such as the FPUT-$\alpha$ or $\beta$ lattices
 ($0.3\lesssim \alpha \lesssim 0.4$).
 
   We emphasize a significance of our model
 from the point of view of future studies.
 The UFL is expected to be a good starting point
 to study the mechanism of emerging of thermal resistance.
 It is possible to gradually introduce the Umklapp processes into the UFL
 by perturbing its potential functions.
 Therefore,
 the mechanism may be clarified
 by numerically observing
 what kind of elementary process is occurring in the perturbed UFL,
 i.e., by identifying the scatterer and scattering process of phonons.
 
   This paper is organized as follows.
 In Sec.~\ref{sec:model},
 we describe the UFL model.
 In Sec.~\ref{sec:simulation},
 we describe details of our numerical simulation of heat transport in the UFL.
 In Sec.~\ref{sec:numerical_results},
 we show numerical results of the simulation.
 Finally, conclusions are drawn in Sec.~\ref{sec:conclusion}.
%
%
%
\section{Umklapp-free lattice model}
\label{sec:model}

   The model we constructed is a class of infinite lattices
 with long-range nonlinear interactions
 which is described by the Hamiltonian 
\begin{eqnarray}
 H &=& \sum_{n=-\infty}^{\infty}\frac{1}{2}\,p_{n}^2
 +\sum_{n=-\infty}^{\infty}
 \left[\,\frac{\mu_0}{2}q_n^2+\frac{\mu_1}{2}(q_{n+1}-q_n)^2\,\right]
\nonumber
\\
 && +~ \beta \sum_{n=-\infty}^{\infty}\sum_{r=1}^{\infty}\frac{1}{4r^2}
 \left\{q_{n+r}-(-1)^rq_n\right\}^4.
\label{eqn:UFL_Hamiltonian}
\end{eqnarray}
 This model corresponds to a one-dimensional chain of unit-mass particles,
 where $n$th particle has its position $x_n=an+q_n$
 given by the lattice spacing constant $a$
 and the relative displacement $q_n$.
 In Eq.~(\ref{eqn:UFL_Hamiltonian}),
 $p_n$ is the momentum of $n$th particle,
 $\mu_0$ and $\mu_1$ are coefficients of
 the harmonic on-site and interaction potentials,
 and $\beta>0$ is the nonlinearity strength.
 Arbitrary non-negative values are possible for $\mu_0$ and $\mu_1$.
 The coupling strength
 between the $r$th neighboring particles
 is proportional to $1/r^2$.
 We call this lattice the UFL.
 Note that
 the UFL should be regarded as only a mathematically idealized model
 since the nonlinear interaction term in Eq.~(\ref{eqn:UFL_Hamiltonian})
 is physically unnatural due to the factor $(-1)^r$.

   The equations of motion derived
 from the Hamiltonian (\ref{eqn:UFL_Hamiltonian})
 are given by
\begin{eqnarray}
 \ddot{q}_n &=& -\mu_0 q_n+\mu_1\left(q_{n+1}-2q_{n}+q_{n-1}\right)
\nonumber
\\
 &+& \beta\sum_{r=1}^{\infty}\frac{1}{r^2}\left[
 \left\{(-1)^r q_{n+r}-q_n\right\}^3-\left\{q_n-(-1)^rq_{n-r}\right\}^3
 \right],~~
\label{eqn:Eqs_motion}
\end{eqnarray}
 where $n\in\mathbb{Z}$.
 Note that the total momentum $\sum_{n=-\infty}^{\infty}p_n$ is not conserved
 by Eq.~(\ref{eqn:Eqs_motion})
 regardless of the value of $\mu_0$,
 as shown in Appendix~\ref{sec:A}.
 
   Define the normal mode coordinates $U(k)$
 via the discrete Fourier transformation
\begin{equation}
 U(k)=
 \frac{1}{\sqrt{2\pi}}\sum_{n=-\infty}^{\infty}q_n e^{-ik n},
\label{eqn:def_U}
\end{equation}
 where we restrict the range of wavenumber $k$ to the first Brillouin zone,
 i.e., $k\in \mathbb{T}\equiv(-\pi,\pi]$.
 If we rewrite Eq.~(\ref{eqn:Eqs_motion}) in terms of $U(k)$,
 we can obtain the equation
\begin{eqnarray}
 \ddot{U}(k)+\nu_k^2U(k)
 &=& \frac{4\beta}{\pi}
 \int_{\mathbb{T}^3}\! dk_1 dk_2 dk_3 \phi_0(k_1,k_2,k_3,k)
\nonumber
\\
 &\times &
 U(k_1)U(k_2)U(k_3) \delta(k_1+k_2+k_3-k),~~
\label{eqn:Eqs_motion_U}
\end{eqnarray}
 where $U(k)$ depends on time $t$,
 $\phi_0$ is a time-independent function of $(k_1,k_2,k_3,k)$,
 $\delta$ is Dirac's delta function,
 and $\nu_k^2$ is given by
\begin{equation}
 \nu_k^2=\mu_0+4\mu_1\sin^2(k/2).
\label{eqn:nu_k}
\end{equation}
 Details of the derivation of Eq.~(\ref{eqn:Eqs_motion_U}) are described
 in Appendix~\ref{sec:B}.

   Ordinary one-dimensional lattices with quartic potentials
 have the mode couplings specified by both $k_1+k_2+k_3-k=0$ and $\pm 2\pi$
 (cf. Appendix~\ref{sec:B1}).
 The former $k_1+k_2+k_3-k=0$ is called the normal process
 while the latter $k_1+k_2+k_3-k=\pm 2\pi$ the Umklapp process.
 Equation~(\ref{eqn:Eqs_motion_U}) shows that
 four normal modes are coupled
 only when their wavenumbers satisfy the condition $k_1+k_2+k_3-k=0$
 while the couplings of $\pm 2\pi$ are not allowed.
 This mode coupling rule is a peculiarity of the UFL,
 and it indicates the non-existence of the Umklapp process.
 As mentioned in Sec.~\ref{sec:introduction},
 the UFL closely relates with the PISL.
 Their relation is described in Appendix~\ref{sec:C}.
%
%
%
\section{Simulation setup}
\label{sec:simulation}
\begin{figure}[t]
\begin{center}
 \includegraphics[width=85mm,height=25mm]{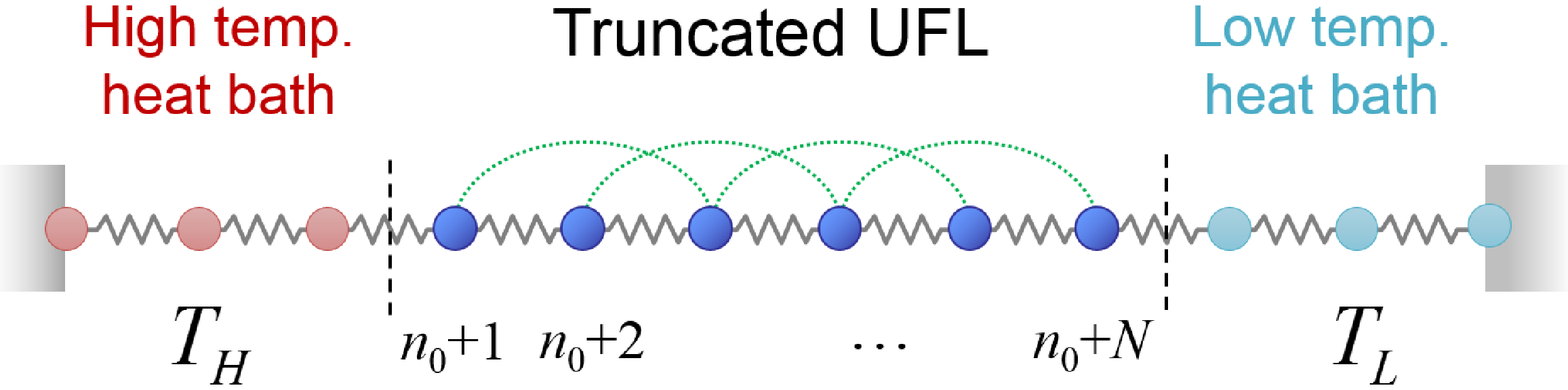}
 \caption{Illustration of simulation model.
 Green dotted line represents the long-range interaction,
 where $M=2$ case is illustrated.}
 \label{fig:sys_config}
\end{center}
\end{figure}

   In order to study the heat transport
 by non-equilibrium molecular dynamics simulation,
 we introduce an approximate version of the UFL
 which has
 truncated long-range interactions
 up to length $M$,
 which we call the {\it truncated UFL}.
 This lattice is not exactly free from the Umklapp processes:
 the mode coupling terms specified by $k_1+k_2+k_3-k=\pm 2\pi$
 appear in its equation of motion
 in the normal mode coordinates $U(k)$.
 Those Umklapp terms become smaller and vanish as $M\to +\infty$.
 Thus,
 the truncated UFL satisfies
 the condition of non-existence of the Umklapp process
 in good approximation when $M$ is large enough.

   In numerical simulations,
 we use a finite-size truncated UFL
 equipped with stochastic Langevin thermostats in its both ends.
 The equations of motion of our simulation model
 are given as follows:
\begin{eqnarray}
 \ddot{q}_n = \lambda\left(q_{n+1}-2q_{n}+q_{n-1}\right)-\gamma\dot{q}_n+\zeta_n(t)
\label{eqn:NEMD_high_eq}
\end{eqnarray}
for $n\in I_H\cup I_L$ and
\begin{eqnarray}
 \ddot{q}_n &=& -\mu_0 q_n+\mu_1\left(q_{n+1}-2q_{n}+q_{n-1}\right)
\nonumber
\\
 &+& \beta\sum_{r=1}^{M}\frac{1}{r^2}\left[
 \left\{(-1)^r q_{n+r}-q_n\right\}^3\!-\left\{q_n-(-1)^rq_{n-r}\right\}^3
 \right]
\label{eqn:NEMD_lattice_eq}
\end{eqnarray}
 for $n\in I$,
 where $I_H=\{1,2,\dots,n_0\}$ and $I_L=\{N+n_0+1,\dots,N+2n_0\}$
 are the sets of indices of particles equipped with
 high and low temperature thermostats, respectively,
 and
 $I=\{n_0+1,\dots,n_0+N\}$ is the set of indices
 for the truncated UFL.
 The constant $\lambda$ can be different from $\mu_1$,
 but we assume $\lambda=\mu_1$ in the present simulation for simplicity.
 The range of nonlinear interactions is truncated up to $M$ in the model.
 In addition,
 the sum in Eq.~(\ref{eqn:NEMD_lattice_eq}) is taken
 only for the terms
 $\{(-1)^r q_{n+r}-q_n\}^3$ satisfying $n+r \le n_0+N$
 and the terms
 $\{q_n-(-1)^rq_{n-r}\}^3$ satisfying $n-r\ge n_0+1$.
 This implies that
 the nearest neighbor harmonic coupling is assumed
 between $n_0$ and $n_0+1$ particles
 and between $n_0+N$ and $n_0+N+1$ particles,
 which are connections
 between the truncated UFL and the heat baths.
 As for the boundary conditions,
 we assume $q_0=q_{N+2n_0+1}=0$.
 Figure~\ref{fig:sys_config} illustrates the simulation model.

   In Eq.~(\ref{eqn:NEMD_high_eq}),
 $-\gamma\dot{q}_n+\zeta_n(t)$ represents the Langevin thermostat,
 where $\gamma>0$ is a constant
 and $\zeta_n(t)$ is the white Gaussian noise
 having the properties
\begin{eqnarray}
 &&\left\langle \zeta_n(t) \right\rangle=0,
\\
 &&\left\langle \zeta_n(t)\zeta_m(s) \right\rangle=2\gamma T\delta_{n,m}\delta(t-s),
\end{eqnarray}
 where $\langle\cdot\rangle$ denotes the averaging
 over realizations of $\zeta_n(t)$,
 $\delta_{n,m}$ is Kronecker's delta,
 and $\delta$ is Dirac's delta function.
 The parameter $T$ represents the thermostat temperature,
 which is set as $T=T_H$ and $T_L$
 for the high and low temperature sides, respectively.

   The heat flux can be measured
 via a simple expression
 at the boundaries of the truncated UFL,
 i.e., $n=n_0+1$ and $n_0+N$ particles.
 Each of these two particles is coupled with
 its nearest neighbor thermostated particle
 via harmonic interaction force only.
 Noting this fact,
 we obtain the expression for the heat flux $J_1$,
 which is the energy transported
 from $n_0$th particle to $(n_0+1)$th one
 per unit time,
 as follows:
\begin{equation}
 J_1 = -\left\langle\,
 \dot{q}_{n_0+1}\cdot\mu_1\left(q_{n_0+1}-q_{n_0}\right)\,
 \right\rangle_{\tau},
\label{eqn:J1}
\end{equation}
 where $\langle\cdot\rangle_{\tau}$ represents
 averaging over a long time $\tau$, i.e.,
 $\langle X \rangle_{\tau}=\tau^{-1}\int_{0}^{\tau}X(t)dt$
 for an arbitrary quantity $X(t)$.
 Similarly,
 we can obtain the heat flux $J_2$
 at $(n_0+N)$th particle as follows:
\begin{equation}
 J_2 = -\left\langle\,
 \dot{q}_{n_0+N}\cdot\mu_1\left(q_{n_0+N+1}-q_{n_0+N}\right)\,
 \right\rangle_{\tau}.
\label{eqn:J2}
\end{equation}
 If we measure the heat flux
 at an inner particle of the truncated UFL with $n\in\{n_0+2,\dots,n_0+N-1\}$,
 we will have a more complex expression of heat flux
 due to the long-range interactions.
 So, we chose the two boundary particles.
 In the simulation,
 we compute the heat flux $J$
 by the average of $J_1$ and $J_2$, i.e.,
\begin{equation}
 J=\frac{1}{2}\left(J_1+J_2\right).
\label{eqn:J}
\end{equation}
 The thermal conductivity $\kappa$ is defined by 
\begin{equation}
 \kappa=\frac{J}{(T_H-T_L)/N}.
\label{eqn:kappa}
\end{equation}
 We will focus on the $N$-dependence of $\kappa$.
 It is well known that
 one-dimensional lattices exhibit
 the power law $\kappa \propto N^\alpha$ with $0 \le \alpha \le 1$
 \cite{Lepri-2003}.
 The heat transport is called {\it normal} when $\alpha=0$
 while it is called {\it anomalous} when $\alpha>0$.
 In particular,
 it is called the {\it ballistic} heat transport when $\alpha=1$,
 and this implies the state of vanishing thermal resistance.

   We introduce spectral energy flux
 to study the heat transport process in detail.
 If we neglect the nonlinear forces and only consider the harmonic one
 in Eq.~(\ref{eqn:Eqs_motion}),
 we can define
 the harmonic part of the total heat flux as follows:
\begin{equation}
 J_{H,tot}=-\frac{\mu_1}{2}\sum_{n=n_0+1}^{n_0+N-1}
 (\dot{q}_{n+1}+\dot{q}_n)\,(q_{n+1}-q_n).
\label{eqn:JH_tot}
\end{equation}
 Let  $u_k \in \mathbb{C},\,k=-N/2+1,\ldots, N/2$
 be the mode amplitudes defined by the transformation
\begin{eqnarray}
 q_{n_0+n}= \frac{1}{\sqrt{N}}\!\sum_{k=-N/2+1}^{N/2}\!
 {u_k \exp\left[{-i\frac{2\pi k}{N}n}\right]},
 ~~n=1,2,\dots, N,~~
 \label{eqn:uk}
\end{eqnarray}
 where $u_{-k}=\bar{u}_k$ holds since $q_{n_0+n}\in\mathbb{R}$.
 $\bar{u}_k$ stands for the complex conjugate of $u_k$.
 In terms of $u_k$,
 we can decompose $J_{H,tot}$ into the form
 $J_{H,tot}=\sum_{k=1}^{N/2-1}J_H(k)$ with
\begin{equation}
 J_H(k)=2\omega_k v_k\,\mathrm{Im}\left[\dot{u}_k\bar{u}_k\right],
 \quad k=0,1,\dots,N/2,
\label{eqn:JH}
\end{equation}
 where $\omega_k$ and $v_k$ are defined by
 $\omega_k=2\sqrt{\mu_1}\,|\sin(\pi k/N)|$
 and $v_k=\sqrt{\mu_1}\,\mathrm{sgn}(k)\cos(\pi k/N)$.
 This quantity $J_H(k)$ is
 the harmonic part of the net energy flux
 carried by two modes with wavenumbers $\pm k$.
 The derivation of Eq.~(\ref{eqn:JH}) is described
 in Appendix~\ref{sec:D}.
%
%
%
\section{Simulation results}
\label{sec:numerical_results}
\begin{figure}[t]
\begin{center}
 \includegraphics[width=75mm,height=58mm]{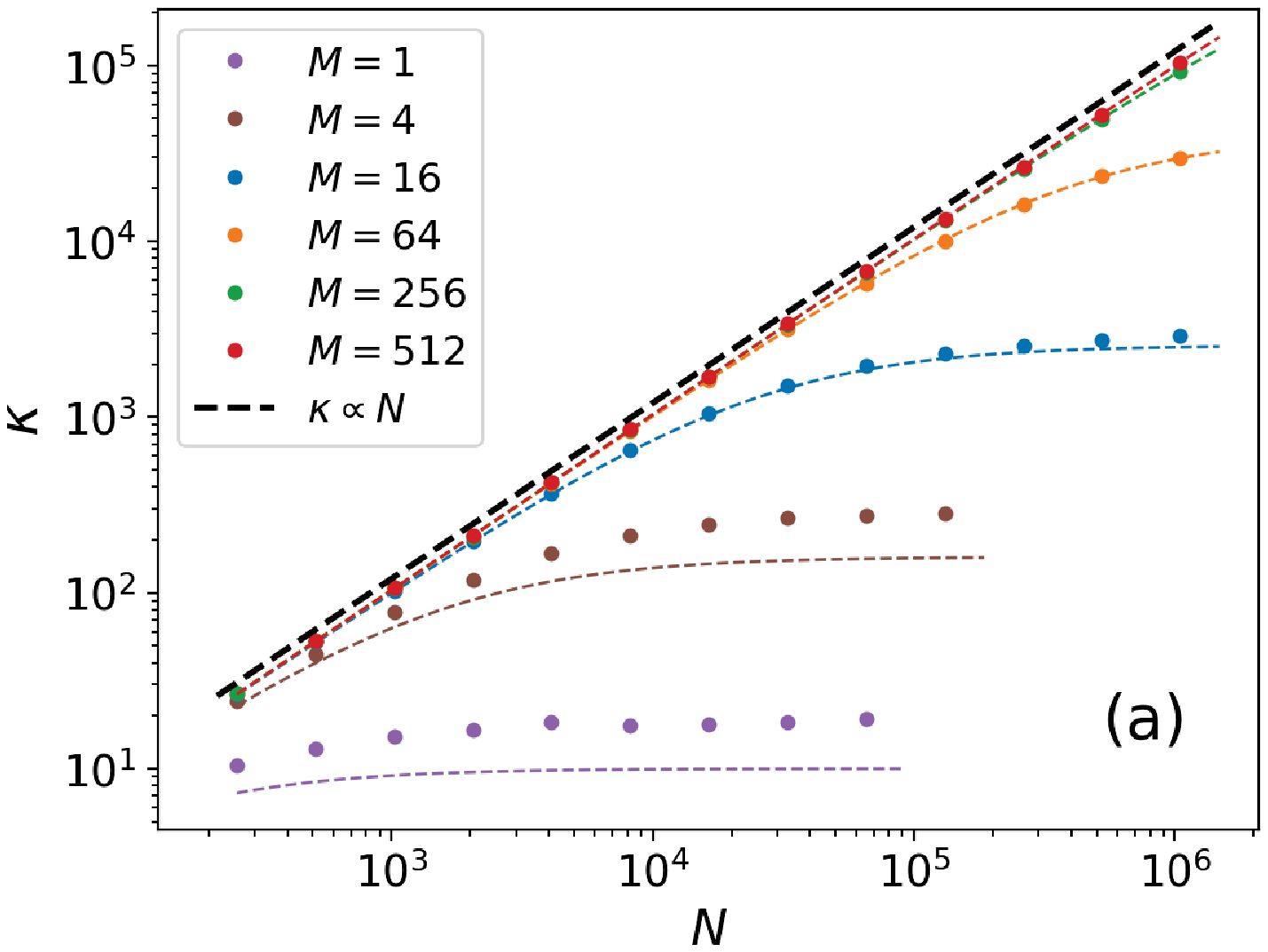}
 \includegraphics[width=75mm,height=58mm]{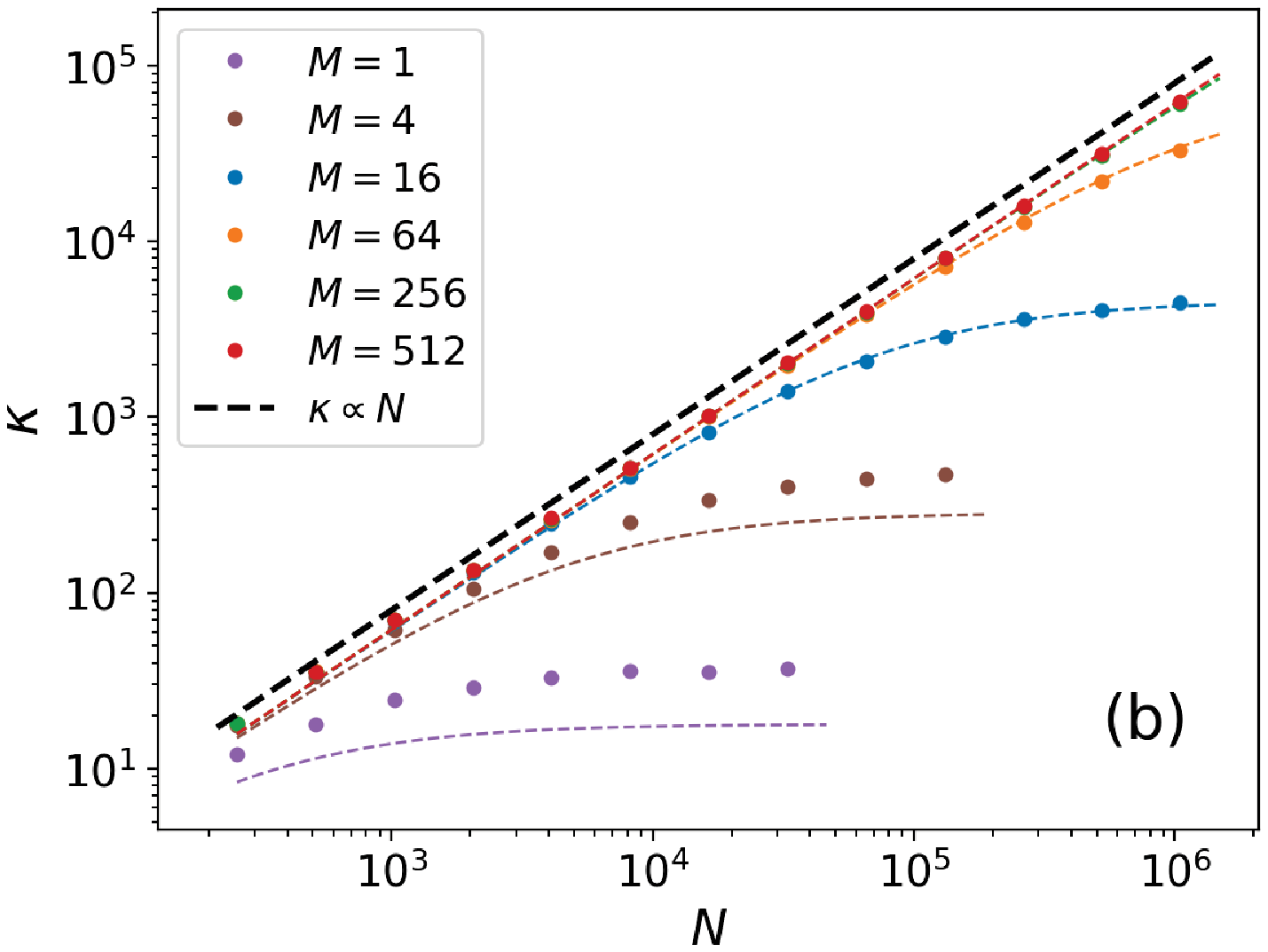}
 \caption{Thermal conductivity $\kappa$ vs. lattice size $N$
 for truncated UFL ($\mu_1=1,\,\beta=0.1$)
 (a) without on-site potential ($\mu_0=0$)
 and (b) with on-site potential ($\mu_0=1$).
 Results are shown by filled circles for different values of $M$.
 Temperatures are $T_H=1.2$ and $T_L=0.8$.
 Black dashed line represents the ballistic power law $\kappa \propto N$.
 Colored dashed lines are the fitting curves
 by experimental formula~(\ref{eqn:exp_formula}).}
 \label{fig:kappa_N}
\end{center}
\end{figure}

   We numerically solved
 Eqs.~(\ref{eqn:NEMD_high_eq}) and (\ref{eqn:NEMD_lattice_eq})
 to compute the thermal conductivity $\kappa$
 for different lattice sizes $N$,
 by using the Verlet scheme with time step $\Delta t=0.05$.
 Computation of the long-range nonlinear interaction forces
 in Eq.~(\ref{eqn:NEMD_lattice_eq}) is time-consuming
 for large values of $M$.
 To overcome this difficulty,
 we utilized GPU (\,NVIDIA GeForce RTX3080\,)
 for high-speed computation. 
 The parameter values used in the simulation are
 $\lambda=1$, $\gamma=0.2$, $T_H=1.2$, $T_L=0.8$, and $n_0=10$.

   Figures~\ref{fig:kappa_N}(a) and \ref{fig:kappa_N}(b) 
 show the logarithmic plots of $\kappa$ as a function of $N$
 for $\mu_0=0$ and $\mu_0=1$ cases,
 i.e, the lattices without and with harmonic on-site potential, respectively.
 The weakly nonlinear case of $\mu_1=1$ and $\beta=0.1$ is assumed.
 The numerical results are shown
 for different values of the coupling length $M$ from $M=1$ to $512$.
 
\begin{figure}[t]
\begin{center}
 \includegraphics[width=70mm,height=50mm]{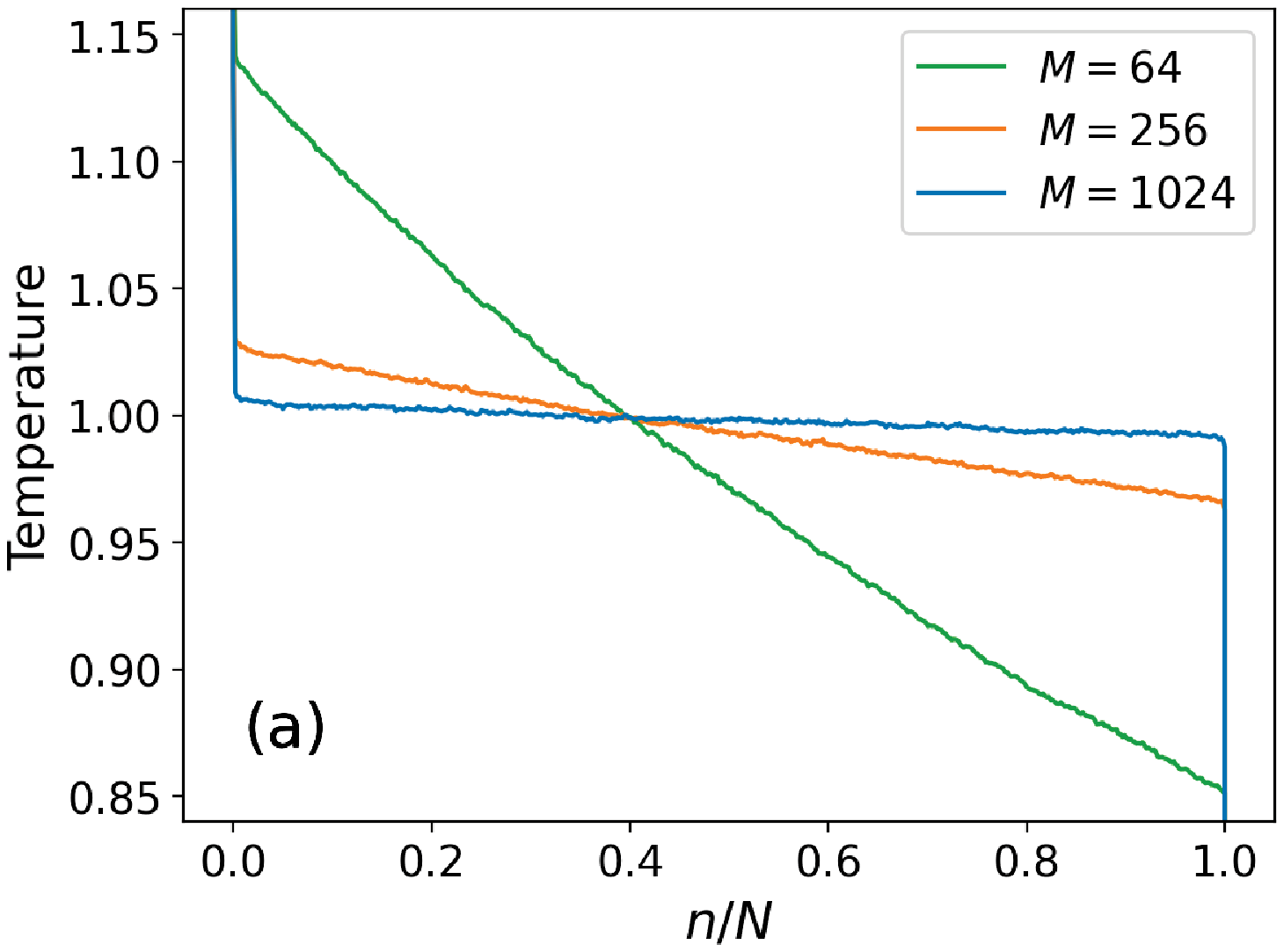}
 \includegraphics[width=70mm,height=50mm]{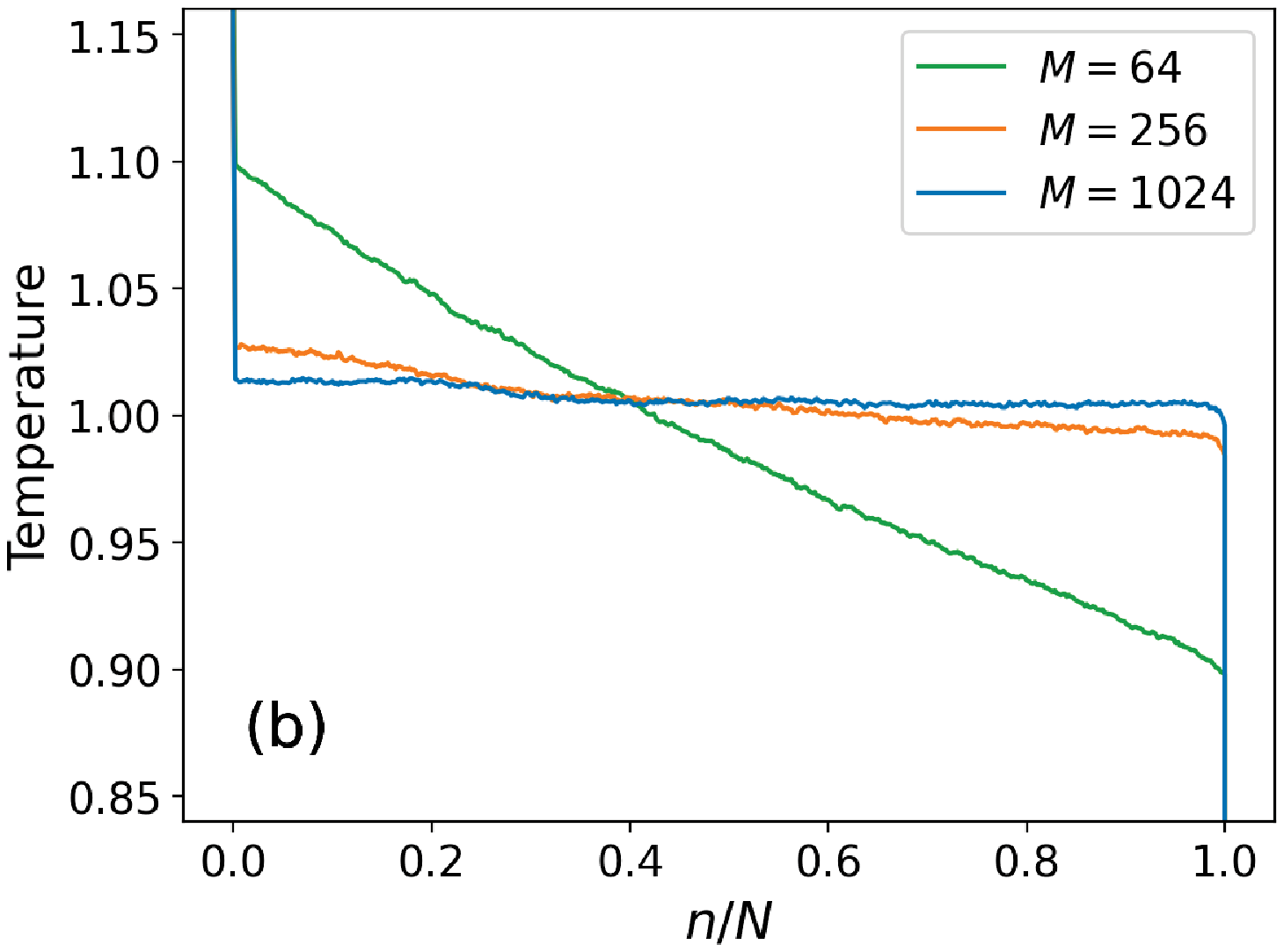}
 \caption{
 Temperature profile plotted vs. $n/N$
 for the truncated UFL ($\mu_1=1,\,\beta=0.1$)
 (a) without on-site potential ($\mu_0=0$)
 and (b) with on-site potential ($\mu_0=1$).
 Profiles are shown for $M=64$, $256$, and $1024$.
 Parameters are $N=2^{20}$, $T_H=1.2$, and $T_L=0.8$.}
 \label{fig:temp}
\end{center}
\end{figure}

   In the simulations,
 we monitored the heat flux $J$ given by Eq.~(\ref{eqn:J})
 as a function of $\tau$,
 which tends to converge to a constant
 as the averaging period $\tau$ increases.
 We used the convergence of $J(\tau)$
 as a criterion for the system
 to have reached a steady state.
 In addition,
 we also monitored convergence of the spatial temperature profile
 (cf. Fig.~\ref{fig:temp}).
 
   In Fig.~\ref{fig:kappa_N}(a),
 the scaling of $\kappa$ with respect to $N$
 precisely coincides with the ballistic one $\kappa \propto N$
 over the whole range of simulation, i.e., up to $N=2^{20}$,
 in the case of $M=512$.
 For smaller values of $M$,
 the scaling is close to $\kappa \propto N$
 as $N$ increases up to a certain value $N_c$,
 but it deviates from $\kappa \propto N$
 as $N$ further increases.
 The values of $N_c$ are approximately found as
 $N_c\simeq 2^{19}$, $2^{14}$, and $2^{11}$
 for $M=256$, $64$, and $16$, respectively.
 $N_c$ decreases as $M$ decreases.
 In Fig.~\ref{fig:kappa_N}(b),
 qualitatively the same behavior of $\kappa$ is observed.
 The ballistic transport is clearly observed for $M=512$ also
 in Fig.~\ref{fig:kappa_N}(b).

   We are interested in
 the asymptotic scaling of $\kappa$ in the limit $N\to +\infty$,
 although numerical results are available only for finite $N$ values.
 An experimental formula is useful to infer the asymptotic scaling,
 and we have found it in the form
\begin{equation}
 \kappa=\frac{aN}{1+bN/M^2},
\label{eqn:exp_formula}
\end{equation}
 where $a$ and $b$ are the fitting parameters
 and their values are obtained as
 $a=0.1032$, $b=0.009904$ for $\mu_0=0$
 and $a=0.06314$, $b=0.004169$ for $\mu_0=1$, respectively.
 In Figs.~\ref{fig:kappa_N}(a) and \ref{fig:kappa_N}(b),
 the curves of Eq.~(\ref{eqn:exp_formula})
 with these values of $(a,b)$ are also shown. 
 A good agreement
 between Eq.~(\ref{eqn:exp_formula}) and the numerical results
 is confirmed in each figure,
 except for the small $M$ cases of $M=1$ and $4$.
 This agreement suggests that
 it is a reasonable experimental formula
 at least for values of $M$ not too small.
 Once we accept Eq.~(\ref{eqn:exp_formula}),
 we can infer the behavior of $\kappa$ in the limit $N\to +\infty$.
 Equation~(\ref{eqn:exp_formula}) indicates that
 the asymptotic scaling $\kappa\propto N$ holds
 if we take the limit $N\to +\infty$
 keeping the ratio $M/N$ as a constant.
 
   As shown in Figs.~\ref{fig:kappa_N}(a) and \ref{fig:kappa_N}(b),
 the truncated UFLs with $M=512$,
 where the non-existence of the Umklapp process holds
 in good approximation,   
 exhibit the ballistic heat transport
 regardless of harmonic on-site potential.
 In contrast,
 in the cases of non-negligible Umklapp process,
 i.e., smaller $M$ cases,
 the ballistic heat transport breaks down for $N>N_c$.
 Based on these numerical observations
 and the inference via Eq.~(\ref{eqn:exp_formula}),
 it may be concluded that
 the thermal resistance is never caused by the normal processes
 but only by the Umklapp one.
 That is, we have validated Peierls's hypothesis.
 We remark that
 it is not clear here whether all the Umklapp processes are resistive
 or only some of them are so.
 
   One might expect the possibility that
 the ballistic transport in the UFL is caused 
 simply by instantaneous energy transport over long distances 
 via the long-range interaction forces.
 This issue has been studied
 for some nonlinear lattices
 with the long-range coupling coefficient $1/r^c$
 \cite{Iubini-2018}.
 It has been shown that
 for $c>1$ such long-distance transport is non-dominant.
 This result is suggestive that
 the ballistic transport in UFL is being caused
 by the lack of Umklapp process.
 
   The ballistic transport observed in the UFL
 is somewhat surprising from the fact that
 the total momentum is not conserved
 by Eq.~(\ref{eqn:Eqs_motion}).
 Table~I summarizes known results
 for the type of heat transport and the total momentum conservation property
 in several one-dimensional nonlinear lattices.
 A common belief is that momentum non-conserving lattices
 belong to the class of normal heat transport,
 and this belief has been corroborated
 in various such lattices
 \cite{Casati-1984,Mimnagh-1997,Prosen-1992,Hu-1998,Hu-2000,Aoki-2000,Cintio-2018}.
 As Table I shows,
 all the momentum non-conserving lattices studied so far
 exhibit the normal heat transport,
 except for an example mentioned below.
 We emphasize that
 the UFL is a counter example against this common belief.
%
\begin{table}[t]
\begin{center}
\begin{tabular}{c|c|c}
\hline
 ~ & Non-conserving & Conserving
 \\
\hline
 \begin{tabular}{c}
 Normal~\\ $\alpha=0$
 \end{tabular} &
 \begin{tabular}{l}
 Ding-a-ling \cite{Casati-1984,Mimnagh-1997} \\
 Ding-dong \cite{Prosen-1992}\\
 Frenkel-Kontrova \cite{Hu-1998} \\
 $\phi^4$ chain \cite{Hu-2000,Aoki-2000} \\
 Toda+harmonic on-site \cite{Cintio-2018} 
 \end{tabular} &
 \begin{tabular}{l}
 Coupled rotators \\ \cite{Giardina-2000}\\
 Modified \\
 ding-a-ling \cite{Lee-2010}
 \end{tabular}  
 \\
\hline
 \begin{tabular}{c}
 Anomalous~\\ $0<\alpha<1$
 \end{tabular} &
 ~ &
 \begin{tabular}{l}
 FPUT-$\alpha$ \cite{Lepri-2000}\\
 FPUT-$\beta$ \cite{Lepri-1998a,Dematteis-2020} \\
 Diatomic Toda \cite{Hatano-1999}
 \end{tabular} \\
\hline
 \begin{tabular}{c}
 Ballistic~\\ $\alpha=1$
 \end{tabular} &
 \begin{tabular}{l}
 UFL 
 \end{tabular} &
 \begin{tabular}{l}
 Toda \cite{Toda-1979}
 \end{tabular}
 \\
\hline
\end{tabular}
\end{center}
\caption{Classification of nonlinear lattice models
 by the type of heat transport
 and the total momentum conservation property.}
\end{table}

   In Table~I,
 we listed only nonlinear lattices of
 the natural Hamiltonian type, i.e., $H=\sum p_n^2/2+V(q_1,\dots,q_N)$.
 Other than this type,
 the ballistic transport has been reported for
 the Izergin-Korepin discrete sine-Gordon model,
 which is an integrable and momentum non-conserving model
 \cite{Prosen-2005}.
 We also mention that
 there is a momentum-conserving coupled map lattice
 which exhibits the normal transport \cite{Giardina-2005}.
 This model was derived
 from a Hamiltonian system with periodic impulsive kicks. 

   The PISL is momentum non-conserving when it has
 a harmonic on-site potential,
 otherwise it is momentum conserving.
 Its heat transport property has been studied
 for the momentum conserving PISL
 in \cite{Bagchi-2017,Yoshimura-2019,Bagchi-2021,Wang-2020}
 while for both types of PISLs in \cite{Yoshimura-2020}.
 Scaling laws close to the ballistic transport,
 i.e., $\kappa\propto N^\alpha$ with $\alpha\simeq 1$,
 are obtained in
 \cite{Bagchi-2017,Yoshimura-2019,Yoshimura-2020,Bagchi-2021},
 whereas a different value $\alpha\simeq 0.71$
 is reported in \cite{Wang-2020}.
 Reasons for this discrepancy in $\alpha$
 are discussed in \cite{Bagchi-2021}.
 At this point,
 a definitive conclusion has not been settled in
 about the value of $\alpha$,
 and
 it is unclear whether the PISLs exhibit
 the ballistic transport or
 non-ballistic but anomalous one.
 So, we did not include the PISL in Table~I.
%
%
\begin{figure}[t]
\begin{center}
 \includegraphics[width=70mm,height=50mm]{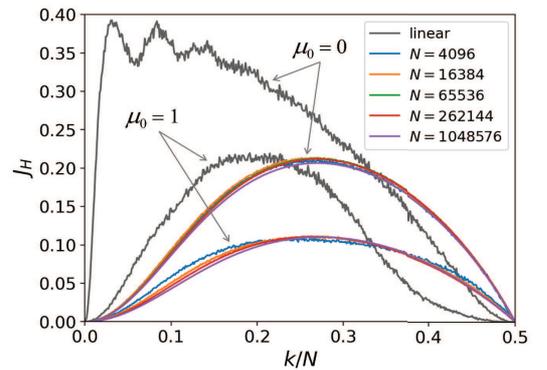}
 \caption{Spectrum of harmonic  energy flux $J_H$ plotted vs. $k/N$.
 Results
 are shown for truncated UFLs ($\mu_1=1,\,\beta=0.1$)
 without on-site potential ($\mu_0=0$)
 and with on-site potential ($\mu_0=1$),
 where $M=512$, $T_H=1.2$, and $T_L=0.8$.
 Results are shown for
 $N=2^{12}$, $2^{14}$, $2^{16}$, $2^{18}$, and $2^{20}$.
 $J_H$ for harmonic lattices ($\mu_1=1,\,\beta=0$)
 with $\mu_0=0$ and $1$ are shown by black line for comparison.}
 \label{fig:flux}
\end{center}
\end{figure}

   Figures~\ref{fig:temp}(a) and \ref{fig:temp}(b) show
 the spatial profile of temperature $T$ as a function of $n/N$
 for $\mu_0=0$ and $1$, respectively,
 where $N=2^{20}$, $n$ is the site number,
 and the local temperature $T$ is defined by
 the time average of kinetic energy, i.e., $T=\langle p_n^2 \rangle_{\tau}$.
 The results are shown for $M=64$, $256$, and $1024$.
 Apart from steep temperature variation in the regions close to the heat baths,
 the temperature gradient becomes smaller as $M$ increases,
 and the flat profile is formed for $M=1024$,
 in each figure.
 This flat profile is one of the characteristics of the ballistic heat transport.

   The harmonic energy flux $J_H$ is plotted against $k/N$
 for $\mu_0=0$ and $1$
 in Fig.~\ref{fig:flux}
 for different $N$,
 respectively, where $M=512$ is fixed.
 The other parameter values are the same as in Fig.~\ref{fig:kappa_N}.
 This figure indicates that
 contribution of the nonlinearity in heat transport
 is substantial since 
 the profiles of $J_H$ are much different
 between the truncated UFLs and the harmonic lattices.
 This fact confirms that
 the ballistic transport observed in Fig.~\ref{fig:kappa_N}
 is not due to a predominance of the linearity. 
 The curves of $J_H(k/N)$ for different $N$ values
 coincide with each other
 in both cases of $\mu_0=0$ and $1$.
 The total amount of $J_H$ over the interval $k/N\in[0,0.5]$,
 which is defined by $\bar{J}_H=\int_{\,0}^{\,0.5}J_H(s)\,ds$ with $s=k/N$,
 is almost independent of $N$.
 This fact is consistent with
 the ballistic scaling $\kappa \propto N$.
 We note that
 $\bar{J}_H$ does not coincides with $J_{H,tot}$:
 they relate with each other as $\bar{J}_H\simeq J_{H,tot}/N$.
 Comparing the profiles of curves of $J_H(k/N)$
 between $\mu_0=0$ and $1$ cases,
 there is a significant difference.
 This fact suggests that
 in our simulation
 the heat transport state is in actual influenced
 by the on-site potential,
 although only similar results are observed
 in Figs.~\ref{fig:kappa_N} and \ref{fig:temp}
 between $\mu_0=0$ and $1$ cases.

   Figure~\ref{fig:flux} shows that
 the normal modes over a broad range of $k/N$,
 especially over an intermediate range from
 $k/N\simeq 0.1$ to $0.45$,
 make non-negligible contributions to the heat transport.
 This shows that
 the heat transport mechanism of the UFL is quite different from
 that of the FPUT lattice, 
 which exhibits the non-ballistic anomalous heat transport.
 In the FPUT lattice,
 only the normal modes with small $k$
 make a dominant contribution
 while $J_H$ is strongly suppressed for the other larger $k$
 as $N$ increases \cite{Dematteis-2020}.
 This suggests that
 those small $k$ modes form solitons
 and
 they induce the anomalous heat transport
 in the FPUT lattice.
 In contrast,
 $J_H$ is small for $k\simeq 0$,
 and this suggests that
 solitons are not formed and 
 the normal modes, i.e., phonons, are the main heat carriers
 in the UFL.
 The heat transport by phonon is the situation
 supposed in Peierls's hypothesis.
\begin{figure}[t]
\begin{center}
 \includegraphics[width=80mm,height=55mm]{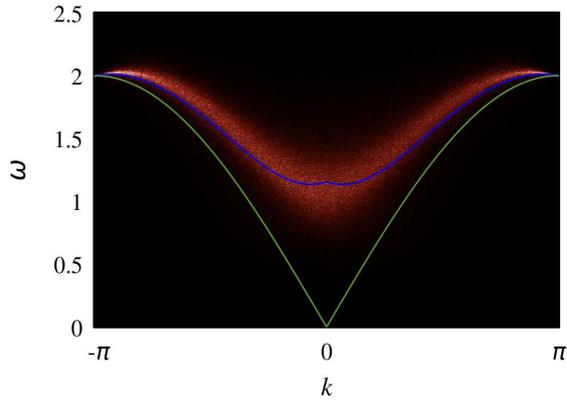} 
 \caption{Space-time Fourier spectrum of
 the steady heat transport state.
 $|S(k,\omega)|^2$ is presented by color. 
 Parameters are $N=4096$, $\mu_0=0$, $\mu_1=1$, and $\beta=0.1$.
 Dispersion curve of the harmonic lattice ($\beta=0$)
 and that of Eq.~(\ref{eqn:dispersion_travel})
 are shown by green and blues lines, respectively.}
 \label{fig:FFT_uk}
\end{center}
\end{figure}

   In order to identify the thermal energy carriers precisely,
 we computed the space-time Fourier spectrum
 defined by
\begin{equation}
 S(k,\omega)=\frac{1}{\sqrt{\tau}}\int_{0}^{\tau}u_k(t)e^{-i\omega t}dt,
\label{eqn:FFT_uk}
\end{equation}
 where $u_k$ is the mode amplitude defined by Eq.~(\ref{eqn:uk})
 and $\tau$ is a time interval taken sufficiently long.
 Figure~\ref{fig:FFT_uk} shows
 the magnitude of $|S(k,\omega)|^2$ by color.
 There clearly appears a narrow strip-like curve
 indicated by bright red color,
 which represents large values,
 above the harmonic dispersion curve.

   Equation~(\ref{eqn:Eqs_motion}) has the traveling wave solutions
\begin{equation}
 q_n(t)=A\cos(kn-\omega t),~~~n\in\mathbb{Z},
\label{eqn:sol_travel} 
\end{equation}
 where $\omega$ depends on $(k,A)$,
 and it is given by the nonlinear dispersion relation
\begin{equation}
 \omega = \sqrt{\,\nu_k^{\,2}+(3/2)\,\pi(\pi-|k|)\,\beta A^2},
\label{eqn:dispersion_travel}
\end{equation}
 where $\nu_k^{\,2}$ is given by Eq.~(\ref{eqn:nu_k}).
 We call this traveling wave the {\it nonlinear phonon}.
 The solution given by Eq.~(\ref{eqn:sol_travel}) is proved
 to be an exact one for $k\in [\pi/3,\pi]$
 in a similar manner to the proof in \cite{Yoshimura-2019},
 while it is an approximate one for $k\in [0,\pi/3)$.

   The curve of Eq.~(\ref{eqn:dispersion_travel})
 fitted to the numerical result by adjusting $A$
 is shown by blue line in Fig.~\ref{fig:FFT_uk},
 and it is in good agreement with the narrow strip-like curve.
 A further numerical evidence is given
 in Appendix~\ref{sec:E}.
 Based on this agreement,
 we may conclude that
 the thermal energy is carried by the nonlinear phonons.
 Moreover,
 figure~\ref{fig:FFT_uk} indicates that
 the nonlinear phonons propagate with subsonic velocities,
 since their maximal group velocity
 $\max_{k\in[0,\pi]}\partial\omega/\partial k$,
 which can be estimated from the curve of Eq.~(\ref{eqn:dispersion_travel})
 in Fig.~\ref{fig:FFT_uk},
 is smaller than the sound velocity $\partial\nu_k/\partial k|_{k=0}$.
 We note that
 each of the nonlinear phonons does not propagate independently,
 but they exchange their energy via the normal processes during propagation,
 because superpositions of Eq.~(\ref{eqn:sol_travel}) are
 no longer exact or approximate solutions of Eq.~(\ref{eqn:Eqs_motion}).
%
%
%
\section{Conclusions}
\label{sec:conclusion}

   We constructed nonlinear lattices
 having a special type of
 long-range quartic interaction potential
 such that the Umklapp process vanishes and only the normal processes exist,
 which we call the UFL.
 It is possible by using the UFL
 to directly verify Peierls's hypothesis that
 only the Umklapp processes can cause the thermal resistance
 while the normal one do not.
 Considering two types of the UFLs
 with and without the harmonic on-site potential,
 we studied their heat transport property
 by non-equilibrium molecular dynamics simulation.
 The numerical results and the experimental formula have shown that
 the ballistic heat transport, i.e., $\kappa\propto N$,
 occurs in the UFLs,
 and justify Peierls's hypothesis.
 Moreover,
 we pointed out the existence of the nonlinear phonons
 and showed that they are the thermal energy carriers
 which propagate with subsonic velocities.
 Finally,
 we emphasize that
 the UFL can be a good starting point
 to study the mechanism of emerging of the thermal resistance
 based on dynamics.
 It may be possible to clarify
 how the thermal resistance emerges via the Umklapp processes
 by perturbing the UFL. 
%
%
%
\section*{Acknowledgment}

   The authors were supported by a
 Grant-in-Aid for Scientific Research (C), No.~19K03654
 from Japan Society for the Promotion of Science (JSPS).
%
%
%
\appendix
%
%
%
\section{Proof of non-conservation of total momentum}
\label{sec:A}

   We show non-conservation of the total momentum in Eq.~(\ref{eqn:Eqs_motion}).
 Precisely speaking,
 the total momentum $\sum_{n=-\infty}^{\infty}p_n$ of the infinite UFL
 does not necessarily have a finite value but may diverge.
 So, we employ its counterpart which is defined by a finite sum.
 Fix an arbitrary $N\in\mathbb{N}$,
 and impose the spatial periodicity condition
 $q_{n+N}=q_n,\,n\in\mathbb{Z}$ to Eq.~(\ref{eqn:Eqs_motion}).
 This is equivalent to considering
 a finite UFL consisting of $N$ particles
 under the periodic boundary condition,
 instead of the infinite UFL.

   Let $M_N=\sum_{n=1}^{N}p_n$. 
 We prove non-conservation of $M_N$.
 It can be checked that
 equation~(\ref{eqn:Eqs_motion}) has
 a solution of the form $q_n(t)=\phi(t),\,n\in\mathbb{Z}$,
 in which all the variables $q_n$ have the same displacement $\phi$.
 This solution apparently satisfies
 the spatial periodicity condition $q_{n+N}=q_n$.
 If we substitute this form into Eq.~(\ref{eqn:Eqs_motion}),
 we have the equation
\begin{equation}
 \ddot{\phi} = -\mu_0\phi-\sigma\phi^3
%
\label{eqn:Eq_phi}
\end{equation}
 where $\sigma=16\beta\sum_{m=1}^{\infty}(2m-1)^{-2}$.
 Equation~(\ref{eqn:Eq_phi}) is regarded as
 that of a Hamiltonian oscillator with the potential
 $V(\phi)=\mu_0\phi^2/2+\sigma\phi^4/4$,
 which is a single-well potential due to $\mu_0\ge 0$ and $\sigma>0$.
 It is clear that
 this equation has a family of non-constant periodic solutions.
 Choose an arbitrary solution from the family.
 Along this solution,
 $\dot{\phi}(t)$ is a non-constant periodic function of $t$.
 This fact implies that
 $M_N=N\dot{\phi}(t)$ is not conserved.
 Thus,
 it has been proved that
 equation~(\ref{eqn:Eqs_motion}) does not conserve the total momentum
 in the sense that $M_N$ is not conserved for any $N\in\mathbb{N}$.
%
%
%
\section{Derivation of equation of motion in normal mode coordinates}
\label{sec:B}

   We describe derivation of Eq.~(\ref{eqn:Eqs_motion_U}) in the main text
 via two steps.
 In the first step,
 we consider a class of lattices
 with general quartic nonlinear interaction potentials,
 and derive its equation of motion in normal mode coordinates.
 In the second step,
 we assume the case of UFL
 and derive Eq.~(\ref{eqn:Eqs_motion_U}).
%
%
\subsection{Normal mode equation for general nonlinear lattices}
\label{sec:B1}

   Consider a class of infinite lattices
 described by the Hamiltonian 
\begin{eqnarray}
 H_{\mathrm{gen}} &=& \sum_{n=-\infty}^{\infty}\frac{1}{2}\,p_{n}^2
 +\sum_{n=-\infty}^{\infty}
 \left[\,\frac{\mu_0}{2}q_n^2+\frac{\mu_1}{2}(q_{n+1}-q_n)^2\,\right]
\nonumber
\\
 && +~ \frac{\beta}{4} \sum_{n=-\infty}^{\infty}\sum_{r=1}^{\infty}b_r
 \left(q_{n+r}-\varepsilon^rq_n\right)^4,
\label{eqn:gen_Hamiltonian}
\end{eqnarray}
 where $\mu_0$ and $\mu_1$ are non-negative constants,
 $\beta>0$ is the nonlinearity strength,
 $b_r$ is the coupling strength between the $r$th neighboring particles,
 and $\varepsilon\in\{-1,1\}$.
 This is a slightly generalized version of Hamiltonian (\ref{eqn:UFL_Hamiltonian}),
 and it describes general nonlinear lattices with quartic two-body interactions.
 For instance,
 Hamitonian (\ref{eqn:gen_Hamiltonian}) describes
 the UFL when $b_r=1/r^2$ and $\varepsilon=-1$,
 while it describes the FPUT-$\beta$ lattice
 when $b_r=\delta_{r,1}$, $\varepsilon=1$, and $\mu_0=0$,
 where $\delta_{r,1}$ is Kronecker's delta.

   The equations of motion derived
 from the Hamiltonian (\ref{eqn:gen_Hamiltonian})
 are given by
\begin{eqnarray}
 \ddot{q}_n &=& -\mu_0 q_n+\mu_1\left(q_{n+1}-2q_{n}+q_{n-1}\right)
\nonumber
\\
 &+& \beta\sum_{r=1}^{\infty}b_r\left[
 \left(\varepsilon^r q_{n+r}-q_n\right)^3-\left(q_n-\varepsilon^rq_{n-r}\right)^3
 \right],~~
\label{eqn:gen_Eqs_motion}
\end{eqnarray}
 where $n\in\mathbb{Z}$.
   
   The normal mode coordinates $U(k)$ are defined by
 the discrete Fourier transformation as follows:
\begin{equation}
 U(k)=\frac{1}{\sqrt{2\pi}}\sum_{n=-\infty}^{\infty}q_n e^{-ik n},
 \quad k\in (-\pi,\pi],
\label{eqn:DFT_q}
\end{equation}
 where $k$ is restricted
 in the first Brillouin zone $\mathbb{T}=(-\pi,\pi]$.
 The inverse transformation is given by
\begin{equation}
 q_n=\frac{1}{\sqrt{2\pi}}\int_{\mathbb{T}}U(k) e^{ik n}dk
 \,, \quad n\in\mathbb{Z}.
\label{eqn:IDFT_U}
\end{equation} 
 Performing the discrete Fourier transformation
 to both sides of Eq.~(\ref{eqn:gen_Eqs_motion}),
 we have
\begin{eqnarray}
\lefteqn{ \ddot{U}(k) + \nu_k^2 U(k)
 = \frac{\beta}{\sqrt{2\pi}}
 \sum_{n=-\infty}^{\infty} e^{-ik n} }
\nonumber
\\
 &&~~\qquad~~ \times
 \sum_{r=1}^{\infty}
  b_r\left[
 \left(\varepsilon^r q_{n+r}-q_n\right)^3-\left(q_n-\varepsilon^r q_{n-r}\right)^3
 \right],~~~~~
\label{eqn:U_eq_no1}
\end{eqnarray}
 where $\nu_k^2=\mu_0+4\mu_1\sin^2(k/2)$.
 Using Eq.~(\ref{eqn:IDFT_U}),
 we have
\begin{eqnarray}
 \varepsilon^r q_{n+r}-q_n
 \!\! &=& \!\! \varepsilon^r\sqrt{\frac{2}{\pi}}\int_{\mathbb{T}}U(k)
 e^{ik n}e^{irk/2}g_r(k)dk,~~
\label{eqn:diff_x+}
\\
 q_n-\varepsilon^r q_{n-r}
 \!\! &=& \!\! \sqrt{\frac{2}{\pi}}\int_{\mathbb{T}}U(k)
 e^{ik n}e^{-irk/2}g_r(k)dk,~~
\label{eqn:diff_x-}
\end{eqnarray}
 where $g_r(k)$ is given by
\begin{equation}
 g_r(k)=\frac{1}{2}\left(e^{irk/2}-\varepsilon^r e^{-irk/2}\right)
 =\left\{
 \begin{array}{ll}
  \cos(rk/2) & \mbox{for odd}~r,
 \\
  i\sin(rk/2) & \mbox{for even}~r.
 \end{array}
 \right.
\label{eqn:g_r}
\end{equation}
 Substituting Eqs.~(\ref{eqn:diff_x+}) and (\ref{eqn:diff_x-})
 into the right hand side of Eq.~(\ref{eqn:U_eq_no1}),
 we have
\begin{eqnarray}
\lefteqn{\ddot{U}(k)+\nu_k^2 U(k)}
\nonumber
\\
 &=& \frac{\beta}{\sqrt{2\pi}}\sum_{n=-\infty}^{\infty} e^{-ik n}
\nonumber
\\
 &&\times \sum_{r=1}^{\infty}
 b_r \left[\,\prod_{j=1}^{3}
 \Biggl\{ \varepsilon^r \sqrt{\frac{2}{\pi}}\int_{\mathbb{T}} U(k_j)
 e^{ik_j n}e^{irk_j/2}g_r(k_j)dk_j \Biggr\}
 \right.
\nonumber
\\
 &&~~ \left.-\prod_{j=1}^{3}
 \Biggl\{\sqrt{\frac{2}{\pi}}\int_{\mathbb{T}} U(k_j)
 e^{ik_j n}e^{-irk_j/2}g_r(k_j)dk_j \Biggr\}
 \,\right]
\nonumber
\\
 &=& \frac{2\beta}{\pi^2}\sum_{n=-\infty}^{\infty}\sum_{r=1}^{\infty}b_r
 \int_{\mathbb{T}^3} dk_1 dk_2 dk_3
 U(k_1)U(k_2)U(k_3)
\nonumber
\\
 &&\times
 \left\{\,\varepsilon^r e^{i(k_1+k_2+k_3-k)r/2}e^{irk/2}
 -e^{-i(k_1+k_2+k_3-k)r/2}e^{-irk/2}\,\right\}
\nonumber
\\
 &&\times~G_r(k_1,k_2,k_3)\,e^{i(k_1+k_2+k_3-k)n},~~~
\label{eqn:U_eq_step1}
\end{eqnarray}
 where $G_r$ is defined by
\begin{equation}
 G_r(k_1,k_2,k_3) = g_r(k_1)g_r(k_2)g_r(k_3).
\label{eqn:G_r} 
\end{equation}
 As for the sum over $n$ in Eq.~(\ref{eqn:U_eq_step1}),
 recall the formula
\begin{equation}
 \sum_{n=-\infty}^{\infty} e^{incx}
 = \frac{2\pi}{c}\sum_{m=-\infty}^{\infty}\delta(x-2\pi m/c),
\label{eqn:FS_delta_a}
\end{equation}
 where $c,\,x\in\mathbb{R}$ are constants.
 We can calculate the sum over $n$ in Eq.~(\ref{eqn:U_eq_step1})
 by applying
 Eq.~(\ref{eqn:FS_delta_a}) with $c=1$
 and $x=k_1+k_2+k_3-k$.
 Then, we obtain
\begin{equation}
 \sum_{n=-\infty}^{\infty}e^{i(k_1+k_2+k_3-k)n}
 =2\pi \sum_{m=-\infty}^{\infty}\delta(k_1+k_2+k_3-k-2\pi m).
\end{equation}
 Using this and denoting $\lambda=k_1+k_2+k_3-k$,
 we can rewrite Eq.~(\ref{eqn:U_eq_step1}) as follows:
\begin{eqnarray}
 \ddot{U}(k)+\nu_k^2 U(k)
 \! &=& \! \frac{4\beta}{\pi}\sum_{r=1}^{\infty}b_r
 \int_{\mathbb{T}^3} dk_1 dk_2 dk_3
 U(k_1)U(k_2)U(k_3)
\nonumber
\\
 &&\times
 \left\{\,\varepsilon^r e^{i(\lambda+k)r/2}-e^{-i(\lambda+k)r/2}\,\right\}
 G_r(k_1,k_2,k_3)
\nonumber
\\
 &&\times
 \sum_{m=-\infty}^{\infty}\delta(\lambda-2\pi m).
\label{eqn:U_eq_step2}
\end{eqnarray}
 Since $-\pi<k_j\le \pi$ and $-\pi\le -k<\pi$,
 we have $-4\pi<\lambda<4\pi$.
 Thus, 
 there are only three possible values of $\lambda$, i.e.,
 $\lambda=0,\pm 2\pi$, which correspond to $m=0,\pm1$, respectively.
 Taking into account this fact,
 we can rewrite Eq.~(\ref{eqn:U_eq_step2}) as
\begin{eqnarray}
 \ddot{U}(k)+\nu_k^2 U(k)
 = \frac{4\beta}{\pi}
 \int_{\mathbb{T}^3} dk_1 dk_2 dk_3
 U(k_1)U(k_2)U(k_3)
\nonumber
\\
 \times \sum_{m=-1}^{1}\!
 \delta(\lambda-2\pi m) \phi_m(k_1,k_2,k_3,k),\,~~
\label{eqn:U_eq_step3}
\end{eqnarray}
 where $\phi_m$ is defined by
\begin{eqnarray}
 \phi_m(k_1,k_2,k_3,k)
 \! &=& \! \sum_{r=1}^{\infty}b_r
 \left\{\,\varepsilon^r e^{i\pi mr}e^{ir k/2}-e^{-i\pi mr}e^{-ir k/2}\,\right\}
\nonumber
\\
 &&\times G_r(k_1,k_2,k_3).
\label{eqn:def_phi_m}
\end{eqnarray}
%
%
%
\subsection{Normal mode equation for UFL}
\label{sec:B2}

   Hereafter,
 we assume the case of UFL, i.e., $b_r=1/r^2$ and $\varepsilon=-1$,
 and derive Eq.~(\ref{eqn:Eqs_motion_U}).
 If we use
 $e^{i\pi mr}=e^{-i\pi mr}$, which follows from $m=0,\pm1$,
 and denote $k'=-k$,
 then we have
\begin{eqnarray}
 \phi_m(k_1,k_2,k_3,-k')
 \! &=& \! \sum_{r=1}^{\infty}\frac{1}{r^2}
 \left\{(-1)^r e^{-ik' r/2}-e^{ik' r/2}\right\}e^{i\pi mr} 
\nonumber
\\
 &&\times G_r(k_1,k_2,k_3).
\label{eqn:def_phi_m'}
\end{eqnarray}
 Let $a=k_1/2$, $b=k_2/2$, $c=k_3/2$, and $d=k'/2$. 
 If we divide the sum in Eq.~(\ref{eqn:def_phi_m'}) into two parts
 and use Eqs.~(\ref{eqn:g_r}) and (\ref{eqn:G_r}),
 then we obtain
\begin{equation}
 \phi_m(k_1,k_2,k_3,-k')
 = K_o(m) + K_e(m),
\label{eqn:phi_m'_Koe}
\end{equation}
 where $K_o(m)$ and $K_e(m)$ are given by
\begin{eqnarray}
 && K_o(m) = -2\sum_{r=\mathrm{odd}}\frac{(-1)^m}{r^2}
 \cos(ra)\cos(rb)\cos(rc)\cos(rd),
\nonumber
\\
\label{eqn:def_K_o}
\\
 && K_e(m) = -2\sum_{r=\mathrm{even}}\frac{1}{r^2}
 \sin(ra)\sin(rb)\sin(rc)\sin(rd).
\nonumber
\\
\label{eqn:def_K_e}
\end{eqnarray}
 The sums in Eqs.~(\ref{eqn:def_K_o}) and (\ref{eqn:def_K_e}) are
 taken over all odd and even $r\in\mathbb{N}$, respectively.

   We want to show
 $\phi_m(k_1,k_2,k_3,-k')=0$
 under the condition $k_1+k_2+k_3+k'=2\pi m$
 for $m=\pm 1$.
 It is easy to see that $\phi_m$ has the property
\begin{equation}
 \phi_m(k_1,k_2,k_3,-k')
 = \phi_{-m}(-k_1,-k_2,-k_3,k').
\end{equation}
 Because of this property,
 if $\phi_m(k_1,k_2,k_3,-k')=0$ holds
 for any $(k_1,k_2,k_3,k')$
 satisfying $k_1+k_2+k_3+k'=2\pi m$,
 then $\phi_{-m}(k_1,k_2,k_3,-k')=0$ also holds
 for any $(k_1,k_2,k_3,k')$
 satisfying $k_1+k_2+k_3+k'=-2\pi m$.
 Thus, it is enough to consider one of the $m=\pm 1$ cases.
 In what follows,
 we show $\phi_1(k_1,k_2,k_3,-k')=0$.

   A simple calculation
 using Eqs.~(\ref{eqn:def_K_o}) and (\ref{eqn:def_K_e})
 leads to
\begin{eqnarray}
 K_o(m)
 &=& -\frac{1}{4}\sum_{r=\mathrm{odd}}\frac{(-1)^m}{r^2}
 \bigl\{\cos(r(a+b+c+d))
\nonumber
\\
 &&+\cos(r(a-b+c+d))+\cos(r(a+b-c+d))
\nonumber
\\
 &&+\cos(r(a+b+c-d))+\cos(r(a+b-c-d))
\nonumber
\\
 &&+\cos(r(a-b+c-d))+\cos(r(a-b-c+d))
\nonumber
\\
 &&+\cos(r(a-b-c-d))\bigr\},
\label{eqn:K_o}
\\
 K_e(m)
 &=& -\frac{1}{4}\sum_{r=\mathrm{even}}\frac{1}{r^2}
 \bigl\{\cos(r(a+b+c+d))
\nonumber
\\
 &&-\cos(r(a-b+c+d))-\cos(r(a+b-c+d))
\nonumber
\\
 &&-\cos(r(a+b+c-d))+\cos(r(a+b-c-d))
\nonumber
\\
 &&+\cos(r(a-b+c-d))+\cos(r(a-b-c+d))
\nonumber
\\
 &&-\cos(r(a-b-c-d))\bigr\}.
\label{eqn:K_e}
\end{eqnarray}
 Assuming $m=1$ and
 substituting Eqs.~(\ref{eqn:K_o}) and (\ref{eqn:K_e})
 into Eq.~(\ref{eqn:phi_m'_Koe}),
 we obtain
\begin{eqnarray}
 && \phi_1(k_1,k_2,k_3,-k')
\nonumber
\\
 &=& \frac{1}{4}\sum_{r=1}^{\infty}\frac{(-1)^{r-1}}{r^2}
 \bigl\{\cos(r(a+b+c+d)) +\cos(r(a+b-c-d))
\nonumber
\\
 &&~~+\cos(r(a-b+c-d))+\cos(r(a-b-c+d))\bigr\} 
\nonumber
\\
 &&+ \frac{1}{4}\sum_{r=1}^{\infty}\frac{1}{r^2}
 \{\cos(r(a-b-c-d)) +\cos(r(a-b+c+d))
\nonumber
\\
 &&~~+\cos(r(a+b-c+d))+\cos(r(a+b+c-d))\bigr\}.
\label{eqn:phi_1}
\end{eqnarray}

   Recall that $a,b,c\in(-\pi/2,\pi/2]$ and $d\in[-\pi/2,\pi/2)$,
 which follow from their definitions.
 We show $\phi_1=0$
 for the wider range $a,b,c,d\in [-\pi/2,\pi/2]$.
 Since $\phi_1$ is invariant for any permutation of $a,b,c,d$,
 we can assume $\pi/2\ge a\ge b\ge c\ge d \ge -\pi/2$.
 Recall that $a+b+c+d=\pi$ holds when $m=1$.
 Note that (i) $a,b>0$ and (ii) $c\ge 0$ have to hold,
 because if $b\le 0$ then $a\ge a+b+c+d=\pi$ and this contradict with $\pi/2\ge a$
 and if $c<0$ then $a+b>a+b+c+d=\pi$ and this contradict with $\pi\ge a+b$.
 In addition, note that the sum of any pair taken from $\{a,b,c,d\}$ is positive,
 i.e., (iii) $x+y\ge 0$ for $x,y\in\{a,b,c,d\}$ such that $x\ne y$,
 because if $x+y< 0$ then $z+w> x+y+z+w=a+b+c+d=\pi$
 being in contradiction with $\pi\ge z+w$,
 where $w$ and $z$ are the elements other than $x$ and $y$.
 
   Noting the properties (i)-(iii),
 we can evaluate the ranges of arguments of cosine functions
 in Eq.~(\ref{eqn:phi_1}) as follows:
\begin{eqnarray*}
 &&\mbox{\it In the first sum;}
\\
 &&~~ a+b+c+d=\pi,
\\
 &&~~ 0\le a+b-c-d=a+b-(c+d)\le a+b\le \pi
\\
 &&~~~~~(\,\because a\ge c,~ b\ge d~;~c+d\ge 0\,),
\\
 &&~~ 0\le a-b+c-d=a+c-(b+d)\le a+c\le \pi
\\
 &&~~~~~(\,\because a\ge b,~ c\ge d~;~b+d\ge 0\,),
\\
 &&~~ -\pi\le -(b+c)\le a-b-c+d 
 \le a+d\le \pi
\\
 &&~~~~~(\,\because a+d\ge 0~;~b+c\ge 0\,),
\\
 &&\mbox{\it In the second sum;}
\\
 &&~~ 0\le -(a-b-c-d)=b+c+d-a=\pi-2a\le \pi
\\
 &&~~~~~(\,\because 0\le 2a\le \pi\,),
\\
 &&~~ 0\le a-b+c+d=a+c+d-b=\pi-2b\le \pi
\\
 &&~~~~~(\,\because 0\le 2b\le \pi\,),
\\
 &&~~ 0\le a+b-c+d=a+b+d-c=\pi-2c\le \pi
\\
 &&~~~~~(\,\because 0\le 2c\le \pi\,),
\\
 &&~~ 0\le a+b+c-d=\pi-2d\le \pi+2|d|\le 2\pi
\\
 &&~~~~~(\,\because -\pi\le 2d\le \pi~\,).
\end{eqnarray*}

   In order to compute the two sums in Eq.~(\ref{eqn:phi_1}),
 recall the following formula
\begin{equation}
 \sum_{r=1}^{\infty}\frac{\cos rx}{r^2}
 =\frac{1}{4}(\varphi(x)-\pi)^2-\frac{\pi^2}{12}, 
\label{eqn:sum_formula}
\end{equation}
 where $\varphi(x)$ is the function given by
\begin{equation}
 \varphi(x)=x-2\pi l \quad \mbox{for}~~x\in (2\pi l,2\pi(l+1)],~~l\in\mathbb{Z}.
\label{eqn:def_varphi}
\end{equation}
 If we replace $x$ with $x+\pi$ in Eq.~(\ref{eqn:sum_formula}),
 we can modify the above formula as follows:
\begin{equation}
 \sum_{r=1}^{\infty}\frac{(-1)^{r-1}}{r^2}\cos rx
 =\frac{\pi^2}{12}-\frac{x^2}{4},\qquad x\in[-\pi,\pi].
\label{eqn:sum_formula2}
\end{equation}
 If we apply Eqs.~(\ref{eqn:sum_formula}) and (\ref{eqn:sum_formula2})
 to Eq.~(\ref{eqn:phi_1})
 with noting the ranges of arguments of cosine functions,
 which were shown above,
 we have
\begin{eqnarray}
 && \phi_1(k_1,k_2,k_3,-k')
\nonumber
\\
 &=& \frac{1}{4}\cdot\Biggl[\,
 \frac{\pi^2}{12}-\frac{\pi^2}{4}
 +\frac{\pi^2}{12}-\frac{(a+b-c-d)^2}{4}
\nonumber
\\
 &&+ \frac{\pi^2}{12}-\frac{(a-b+c-d)^2}{4}
 + \frac{\pi^2}{12}-\frac{(a-b-c+d)^2}{4}
\nonumber
\\
 &&+ \frac{\left\{(b+c+d-a)-\pi\right\}^2}{4}-\frac{\pi^2}{12}
\nonumber
\\
 &&+ \frac{\left\{(a-b+c+d)-\pi\right\}^2}{4}-\frac{\pi^2}{12}
\nonumber
\\
 &&+ \frac{\left\{(a+b-c+d)-\pi\right\}^2}{4}-\frac{\pi^2}{12}
\nonumber
\\
 &&+ \frac{\left\{(a+b+c-d)-\pi\right\}^2}{4}-\frac{\pi^2}{12} \,\Biggr]
\nonumber
\\
 &=& \frac{1}{16}(a+b+c+d)^2-\frac{\pi}{4}(a+b+c+d)+\frac{3}{16}\pi^2
\nonumber
\\
 &=& 0, 
\label{eqn:phi_1=0}
\end{eqnarray}
 where we used $a+b+c+d=\pi$.
 Since it has been proved that
 $\phi_{\pm 1}(k_1,k_2,k_3,-k')=0$
 when $k_1+k_2+k_3+k'=\pm 2\pi$,
 Equation~(\ref{eqn:U_eq_step3}) reduces to
\begin{eqnarray}
 \ddot{U}(k)+\nu_k^2 U(k)
 &=& \frac{4\beta}{\pi}
 \int_{\mathbb{T}^3}\! dk_1 dk_2 dk_3
 \phi_0(k_1,k_2,k_3,k)
\nonumber
\\
 &\times & U(k_1)U(k_2)U(k_3)
 \delta(k_1+k_2+k_3-k).
\nonumber
\\
\label{eqn:U_eq_final}
\end{eqnarray} 
 This equals to Eq.~(\ref{eqn:Eqs_motion_U}).
%
%
%
\section{Relation between UFL and PISL}
\label{sec:C}

   The PISL was originally constructed as
 a finite-size lattice with the periodic boundary condition
 \cite{Doi-2016,Doi-2020}.
 Its extension to the infinite-size one is described by the Hamiltonian
\begin{eqnarray}
 H &=& \sum_{n=-\infty}^{\infty}\frac{1}{2}\,p_{n}^2
 +\sum_{n=-\infty}^{\infty}
 \left[\,\frac{\mu_0}{2}q_n^2+\frac{\mu_1}{2}(q_{n+1}-q_n)^2\,\right]
\nonumber
\\
 && +~ \beta \sum_{n=-\infty}^{\infty}\sum_{r=1}^{\infty}\frac{1}{4r^2}
 \left(q_{n+r}-q_n\right)^4,
\label{eqn:PISL_Hamiltonian}
\end{eqnarray}
 which corresponds to the case of $b_r=1/r^2$ and $\varepsilon=1$
 in Eq.~(\ref{eqn:gen_Hamiltonian}).
 The equations of motion are given by
\begin{eqnarray}
 \ddot{q}_n &=& -\mu_0 q_n+\mu_1\left(q_{n+1}-2q_{n}+q_{n-1}\right)
\nonumber
\\
 && + ~\beta\sum_{r=1}^{\infty}\frac{1}{r^2}\left[
 \left(q_{n+r}-q_n\right)^3-\left(q_n-q_{n-r}\right)^3
 \right],
\label{eqn:PISL_Eqs_motion}
\end{eqnarray}
 where $n\in\mathbb{Z}$.
 If we fix an arbitrary even $N\in\mathbb{N}$
 and impose the periodicity condition
 $q_{n+mN}=q_n,\,m\in\mathbb{Z}$
 in Eq.~(\ref{eqn:PISL_Eqs_motion}),
 it reduces to the equations of motion of
 the finite-size periodic PISL in \cite{Doi-2016,Doi-2020}.

   Equations~(\ref{eqn:UFL_Hamiltonian}) and (\ref{eqn:PISL_Hamiltonian})
 show that
 nonlinear potentials of the UFL and the PISL
 are transformed to each other
 by the staggered transformation $q_n \to (-1)^n q_n$.
 In this sense,
 these two lattices correspond to each other.
 This correspondence implies that
 the two lattices have essentially the same dynamics
 when $\mu_0=\mu_1=0$, i.e., the homogeneous potential case.
 
   Let us define $\tilde{U}(m)$ via the transformation
\begin{equation}
 \tilde{U}(m) =
 \frac{1}{\sqrt{2\pi}}\sum_{n=-\infty}^{\infty}(-1)^nq_n e^{-im n},
 \quad m\in (-\pi,\pi],
\label{eqn:def_Um}
\end{equation}
 which is a composition of
 the staggered transformation and
 the discrete Fourier transformation defined by Eq.~(\ref{eqn:DFT_q}).
 If we perform the above transformation
 to rewrite Eq.~(\ref{eqn:PISL_Eqs_motion}) in terms of $\tilde{U}(m)$,
 its nonlinear force part can be computed
 in the same manner as in the UFL.
 Noting the linear force part,
 we obtain the equation
\begin{eqnarray}
 \ddot{\tilde{U}}(m)+\tilde{\nu}_m^2\tilde{U}(m)
 \! &=& \! \frac{4\beta}{\pi}
 \int_{\mathbb{T}^3}\! dm_1 dm_2 dm_3 \phi_0(m_1,m_2,m_3,m)
\nonumber
\\
 \! &\times & \!
 \tilde{U}(m_1)\tilde{U}(m_2)\tilde{U}(m_3) \delta(m_1+m_2+m_3-m),
\nonumber
\\
\label{eqn:Eqs_motion_Um}
\end{eqnarray}
 where
 $\phi_0$ is a time-independent function of $(m_1,m_2,m_3,m)$
 and $\tilde{\nu}_m^2$ is given by
\begin{equation}
 \tilde{\nu}_m^2=\mu_0+4\mu_1\cos^2(m/2).
\label{eqn:nu_m}
\end{equation}
 Equation~(\ref{eqn:Eqs_motion_Um}) has the same form as
 Eq.~(\ref{eqn:Eqs_motion_U}),
 but note that
 the dependence of $\tilde{\nu}_m^2$ on $m$
 is different form that of $\nu_k^2$ on $k$.
 In the non-homogeneous potential case,
 the UFL and the PISL have different dynamics.
 
   Equation~(\ref{eqn:Eqs_motion_Um}) shows that
 four normal modes are coupled
 only when their wavenumbers satisfy the condition $m_1+m_2+m_3-m=0$
 while the couplings of $\pm 2\pi$ are not allowed.
 This mode coupling rule is a peculiarity of the PISL,
 which is similar to that of the UFL.
%
%
%
%
\section{Derivation of the spectral energy flux formula}
\label{sec:D}

   Consider the total harmonic heat flux $J_{H,tot}$ given by Eq.~(\ref{eqn:JH_tot}).
 We approximate $J_{H,tot}$ as follows:
\begin{equation}
 J_{H,tot}\simeq -\frac{\mu_1}{2}\sum_{n=n_0+1}^{n_0+N}
 (\dot{q}_{n+1}+\dot{q}_n)\,(q_{n+1}-q_n).
\label{eqn:JH_tot_approx}
\end{equation}
 where $q_{n_0+N+1}=q_{n_0+1}$.
 The sum is taken only up to $n=n_0+N-1$ in the definition of $J_{H,tot}$.
 In this approximation,
 we added the last term
 $-\mu_1(\dot{q}_{n_0+1}+\dot{q}_{n_0+N})\,(q_{n_0+1}-q_{n_0+N})/2$.
 Since we assume large values of $N$ in our simulation,
 this last term is much smaller than
 the sum of other terms in Eq.~(\ref{eqn:JH_tot_approx}),
 and we can expect Eq.~(\ref{eqn:JH_tot_approx})
 to be a good approximation.
 
   If we substitute Eq.~(\ref{eqn:uk}) into Eq.~(\ref{eqn:JH_tot_approx}),
 we obtain
\begin{eqnarray*}
 &&J_{H,tot}
\nonumber
\\
 &=& \! -\frac{\mu_1}{2N}\sum_{n=1}^{N}
 \Biggl\{\,\sum_{m=-N/2+1}^{N/2} \!\!
 \dot{u}_m\left(e^{-i\theta m}+1\right)e^{-i\theta mn}\,\Biggr\}
\nonumber
\\
 &&\times
 \Biggl\{\,\sum_{k=-N/2+1}^{N/2} \!\!
 u_k\left(e^{-i\theta k}-1\right)e^{-i\theta kn}\,\Biggr\}
\\
 &=& \! -\frac{\mu_1}{2N}\! \sum_{k,m=-N/2+1}^{N/2} \!
 \left[\,
 \sum_{n=1}^{N}\dot{u}_m u_k
 \left(e^{-i\theta m}+1\right)\left(e^{-i\theta k}-1\right)e^{-i\theta(k+m)n}
 \,\right],
\end{eqnarray*}
 where $\theta=2\pi/N$.
 Calculating the sum with respect to $n$,
 we have
\begin{eqnarray}
 J_{H,tot}
 &=&  -\frac{\mu_1}{2N} \! \sum_{k,m=-N/2+1}^{N/2}
 \biggl[\,
 \dot{u}_m u_k
 \left(e^{-i\theta m}+1\right)\left(e^{-i\theta k}-1\right)
\nonumber
\\
 &&\times~ N\left(\delta_{k+m,0}+\delta_{k+m,N}\right)
 \,\biggr],
\label{eqn:JH_tot_step1}
\end{eqnarray}
 where $\delta_{k+m,0}$ and $\delta_{k+m,N}$ are Kronecker's delta.
 The condition $k+m=N$ holds only for $k=m=N/2$,
 and we have $e^{-i\theta m}+1=0$ in this case.
 If we calculate the sum with respect to $m$ in Eq.~(\ref{eqn:JH_tot_step1}),
 taking account of this fact,
 then we have
\begin{eqnarray}
 J_{H,tot}
 &=&  -\frac{\mu_1}{2} \! \sum_{k=-N/2+1}^{N/2-1} \!\!
 u_k\dot{u}_{-k}
 \left(e^{i\theta k}+1\right)\left(e^{-i\theta k}-1\right)
\nonumber
\\
 &=& i\mu_1 \!\! \sum_{k=-N/2+1}^{N/2-1} \!\!
 u_k\dot{u}_{-k}\cdot 2\sin\frac{\pi k}{N}\cos\frac{\pi k}{N}
\nonumber
\\
 &=& i\sum_{k=-N/2+1}^{N/2-1} \!\! \omega_k v_k u_k\dot{u}_{-k},
\label{eqn:JH_tot_step2}
\end{eqnarray}
 where
 $\omega_k=2\sqrt{\mu_1}\,|\sin(\pi k/N)|$
 and $v_k=\sqrt{\mu_1}\mathrm{sgn}(k)\cos(\pi k/N)$.
 Note that
 the term for $k=0$ vanishes due to $\omega_0=0$.
 Dividing the sum in Eq.~(\ref{eqn:JH_tot_step2}) into two parts,
 we can rewrite $J_{H,tot}$ as follows:
\begin{eqnarray}
 J_{H,tot}
 &=& i\sum_{k=1}^{N/2-1} \! \omega_k v_k u_k\dot{u}_{-k}
 + i\sum_{k=1}^{N/2-1} \!\! \omega_{-k} v_{-k} u_{-k}\dot{u}_k
\nonumber
\\
 &=& \frac{1}{i}\sum_{k=1}^{N/2-1} \!
 \omega_k v_k \left(\dot{u}_ku_{-k}-u_k\dot{u}_{-k}\right)
\nonumber
\\
 &=& \sum_{k=1}^{N/2-1}~
 2\,\omega_k v_k\,\mathrm{Im}\left[\dot{u}_k\bar{u}_k\right]
\nonumber
\\
 &=& \sum_{k=1}^{N/2-1}J_H(k),
\label{eqn:JH_tot_step3}
\end{eqnarray}
 where we used $\omega_{-k}v_{-k}=-\omega_k v_k$
 and $u_{-k}=\bar{u}_k$.
%
%
%
\section{Numerical evidence for nonlinear phonons}
\label{sec:E}

    In Fig.~5, 
 the curve of $|S(k,\omega)|^2$ is not a sharp line
 but exhibits non-small line width.
 In addition,
 there is slight discrepancy between
 the dispersion curve of Eq.~(\ref{eqn:dispersion_travel})
 and
 the average profile of $|S(k,\omega)|^2$ curve,
 i.e., the middle line of its strip-like curve.
 Due to these facts,
 it might not be a fully convincing scenario that
 the nonlinear phonons emerge and carry the thermal energy
 in steady heat transport state.
 In this section,
 we give an additional numerical evidence
 to ensure this scenario.

   We have assumed that
 the amplitude $A$ of nonlinear phonons is a constant independent of $k$
 when fitting Eq.~(\ref{eqn:dispersion_travel}) to
 the numerical result of $|S(k,\omega)|^2$.
 However, in actual,
 it may be expected that
 the value of $A$ fluctuates in time
 and moreover the temporal average of $A$ depends on $k$,
 provided that the nonlinear phonons emerge.
 If we take into account these points,
 we may write the amplitude $A$ in the form
\begin{equation}
 A=A_0+\delta A(k)+\varepsilon(k,t),
\label{eqn:A_form}
\end{equation}
 where $A_0$ is a constant,
 $A_0+\delta A(k)$ represents the temporal average of $A$ for $k$,
 and
 $\varepsilon(k,t)$
 represents the temporal fluctuation in $A$ for a given $k$.
 This form may explain
 differences between the numerical result of $|S(k,\omega)|^2$
 and the dispersion curve of Eq.~(\ref{eqn:dispersion_travel}):
 $\varepsilon(k,t)$ causes the line width
 and
 $\delta A(k)$ causes a deviation of 
 the average profile from Eq.~(\ref{eqn:dispersion_travel})
 under substitution of Eq.~(\ref{eqn:A_form}) into Eq.~(\ref{eqn:dispersion_travel}).

\begin{figure}[t]
\begin{center}
\vspace*{3mm}
 \includegraphics[width=80mm,height=55mm]{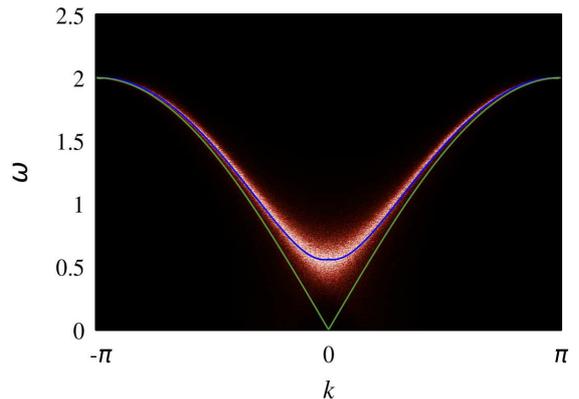} 
 \caption{Space-time Fourier spectrum
 of the steady heat transporting state
 in the UFL of weak nonlinearity $\beta=0.01$.
 $|S(k,\omega)|^2$ is presented by color. 
 Parameters are $N=4096$, $\mu_0=0$, and $\mu_1=1$.
 Dispersion curve of the harmonic lattice ($\beta=0$)
 and that of Eq.~(\ref{eqn:dispersion_travel})
 are shown by green and blues lines, respectively.}
 \label{fig:FFT_uk_b001}
\end{center}
\end{figure}

   The $A$-dependent term is in proportion to $\beta$ in Eq.~(\ref{eqn:dispersion_travel}).
 This fact suggests that
 the influences of $\delta A(k)$ and $\varepsilon(k,t)$
 are small for small values of $\beta$,
 resulting in a better agreement
 of $|S(k,\omega)|^2$ with Eq.~(\ref{eqn:dispersion_travel}) for a $k$-independent constant $A$.
 Figure~\ref{fig:FFT_uk_b001} shows $|S(k,\omega)|^2$
 computed for the UFL of weak nonlinearity $\beta=0.01$,
 where the other parameters are the same as in Fig.~5.
 An excellent agreement is clearly observed.
 This agreement validates the above mentioned scenario.
%
%
%
%

%
%
%

\begin{thebibliography}{9}
 \bibitem{Chang-2008}
  C.~W.~Chang, D.~Okawa, H.~Garcia, A.~Majumdar, and A.~Zettl,
  Phys. Rev. Lett. {\bf 101}, 075903 (2008).
  
 \bibitem{Kittel-2004}
  C.~Kittel,
  {\it Introduction to Solid State Physics} (Wiley, 2004).

 \bibitem{Ashcroft-1976}
  N.~Ashcroft and N.~Mermin,
  {\it Solid State Physics} (Saunders College Publishing, New York, 1976). 
 
 \bibitem{Peierls-1955}
  R.~E.~Peierls, 
  {\it Quantum Theory of Solids} (Oxford University Press, London, 1955).

 \bibitem{Peierls-1997}
  R.~E.~Peierls, On the Kinetic Theory of Thermal Conduction in Crystals,
  in {\it Selected Scientific Papers of Sir Rudolf Peierls},
  edited by R. H. Dalitz and R. Peierls
  (World Scientific, Singapore, 1997).
  
 \bibitem{Jackson-1978}
  E.~A.~Jackson,
  Rocky Mount. J. Math. {\bf 8}, 127 (1978).

 \bibitem{Lepri-2003}
  S.~Lepri, R.~Livi, and A.~Politi,
  Phys. Rep. {\bf 377} 1 (2003). 
  
 \bibitem{Doi-2016}
  Y.~Doi and K.~Yoshimura,
  Phys. Rev. Lett. {\bf 117}, 014101 (2016).

 \bibitem{Doi-2020}
  Y.~Doi and K.~Yoshimura,
  Nonlinearity {\bf 33}, 5142 (2020). 
%
%
 \bibitem{Takeno-1988}
  A.~J.~Sievers and S.~Takeno,
  Phys. Rev. Lett. {\bf 61}, 970 (1988).

 \bibitem{Page-1990}
  J.~B.~Page,
  Phys. Rev. B {\bf 41} 7835 (1990).
  
 \bibitem{Yoshimura-2021}
  K.~Yoshimura and Y.~Doi,
  J. Diff. Eq. {\bf 298}, 560 (2021).

 \bibitem{Yoshimura-2014}
  K.~Yoshimura, Y.~Doi, and M.~Kimura,
  Localized modes in nonlinear discrete systems,
 in {\it Progress in Nanophotonics III}, edited by M. Ohtsu and T. Yatsui
 (Springer, New York, 2014), p.~119.  
%
%
 \bibitem{Bagchi-2017}
  D.~Bagchi,
  Phys. Rev. E {\bf 95}, 032102 (2017).

 \bibitem{Yoshimura-2019}
  K.~Yoshimura and Y.~Doi,
  Proc. of the 2019 Int. Symp. on
  Nonlinear Theory and Its Applications, 399 (2019).
 
 \bibitem{Yoshimura-2020}
  K.~Yoshimura, Y.~Doi, and M.~Ebisu,
  Proc. of the 2020 Int. Symp. on
  Nonlinear Theory and Its Applications, 181 (2020).
  
 \bibitem{Bagchi-2021}
  D.~Bagchi,
  Phys. Rev. E {\bf 104}, 054108 (2021). 
  
 \bibitem{Wang-2020}
  J.~Wang, S.~V.~Dmitriev, and D.~Xiong,
  Phys. Rev. Research {\bf 2}, 013179 (2020).

 \bibitem{Iubini-2018}
  S.~Iubini, P.~Di~Cintio, S.~Lepri, R.~Livi, and L.~Casetti,
  Phys. Rev. E {\bf 97}, 032102 (2018).
%
%
%
 \bibitem{Casati-1984}
  G.~Casati, J.~Ford, F.~Vivaldi, and W.~M.~Visscher,
  Phys. Rev. Lett. {\bf 52}, 1861 (1984).
  
 \bibitem{Mimnagh-1997}
  D.~J.~R.~Mimnagh and L.~E.~Ballentine,
  Phys. Rev. E {\bf 56}, 5332 (1997).
%
 \bibitem{Prosen-1992}
  T.~Prosen and M.~Robnik,
  J. Phys. A {\bf 25}, 3449 (1992).
%
 \bibitem{Hu-1998}
  B.~Hu, B.~Li and H.~Zhao,
  Phys. Rev. E {\bf 57}, 2992 (1998).
%
 \bibitem{Hu-2000}
  B.~Hu, B.~Li, H.~Zhao,
  Phys. Rev. E {\bf 61}, 3828 (2000).

 \bibitem{Aoki-2000}
  K. Aoki and D. Kusnezov,
  Phys. Lett. A {\bf 265}, 250 (2000).
%
 \bibitem{Cintio-2018}
  P.~D.~Cintio, S.~Iubini, S.~Lepri, and R.~Livi,
  Chaos, Solitons and Fractals {\bf 117}, 249 (2018).
%
 \bibitem{Giardina-2000}
  C.~Giardin\'a, R.~Livi, A.~Politi, and M.~Vassalli,
  Phys. Rev. Lett. {\bf 84}, 2144 (2000).
%
 \bibitem{Lee-2010}
  G.~R.~Lee-Dadswell, E.~Turner, J.~Ettinger, and M.~Moy,
  Phys. Rev. E {\bf 82}, 061118 (2010).
%
 \bibitem{Lepri-2000}
  S.~Lepri,
  European Physical Journal B {\bf 18}, 441 (2000).
%
 \bibitem{Lepri-1998a}
  S.~Lepri, R.~Livi, A.~Politi,
  Physica D {\bf 119}, 140 (1998).

 \bibitem{Dematteis-2020}
  G.~Dematteis, L.~Rondoni, D.~Proment, F.~De~Vita, and M.~Onorato,
  Phys. Rev. Lett. {\bf 125}, 024101 (2020).
%
 \bibitem{Hatano-1999}
  T.~Hatano,
  Phys. Rev. E {\bf 59}, R1 (1999).
%
 \bibitem{Toda-1979}
  M.~Toda,
  Physica Scripta {\bf 20}, 424 (1979).
%
 \bibitem{Prosen-2005}
  T.~Prosen and D.~K.~Campbell,
  Chaos {\bf 15}, 015117 (2005).
%
 \bibitem{Giardina-2005}
  C.~Giardin\`a and J.~Kurchan,
  J. Stat. Mech. {\bf 2005}, P05009 (2005).
\end{thebibliography}
\end{document}